\documentclass[%
superscriptaddress,
 amsmath,amssymb,
 aps, physrev,
floatfix,
twocolumn,
]{revtex4-2}

\usepackage{graphicx}% Include figure files
\usepackage{dcolumn}% Align table columns on decimal point
\usepackage{bm}% bold math
\usepackage{hyperref}% add hypertext capabilities
\usepackage{makecell}
\usepackage{enumitem}
\hypersetup{
    colorlinks=true,
    linkcolor=blue,
    citecolor=blue,
    urlcolor=black,
}
\preprint{APS/123-QED}

\usepackage{subfigure,dcolumn}
\usepackage[T2A,T1]{fontenc}    %字体
\usepackage[russian,english]{babel}
\usepackage{color}     %颜色
\usepackage{mathrsfs}  %花体
\usepackage{listings}
\usepackage[marginal]{footmisc}
\usepackage{natbib} %add
\usepackage{xfrac}  %add
\newcommand{\citeptherein}[1]{\citep[and references therein]{#1}}

\begin{document}
\title{On the  High Excitation 7$\alpha$ De-exciting States in $^{28}$Si$^*$}

\author{J. B. Natowitz}\email{Corresponding author. Email: Joenatowitz@gmail.com}
\affiliation{Cyclotron Institute, Texas A$\&$M University, College Station, Texas, USA}

\author{X. G. Cao}
\affiliation{Shanghai Advanced Research Institute, Chinese Academy of Sciences, Shanghai 201210, China}
\affiliation{Shanghai Institute of Applied Physics, Chinese Academy of Sciences, Shanghai 201800, China}
\affiliation{University of Chinese Academy of Sciences, Beijing 101408, China}

\author{A. Bonasera}
\affiliation{Cyclotron Institute, Texas A$\&$M University, College Station, Texas, USA}
\affiliation{Laboratori Nazionali del Sud, INFN, Catania 95123, Italy}

\author{T. Depastas}
\affiliation{Cyclotron Institute, Texas A$\&$M University, College Station, Texas, USA}

\begin{abstract}
A direct comparison and analysis of published spectra for the 7$\alpha$ disassembly of $^{28}$Si projectiles excited in collisions with $^{12}$C at 35 MeV/u reveals significant agreement in the derived excitation energies of high excitation energy resonances observed in two different experiments, in contrast to some earlier conclusions reported in the literature. Many of the observed resonances have excitation energies consistent with those arrived at in recent theoretical investigations explicitly predicting the excitation energies and spins of toroidal nuclei.
An AI-assisted application of well-established statistical filtering techniques reveals identical structures in all spectra investigated. Some additional peaks are observed. The possibility that they correspond to other favored geometries is discussed.

\end{abstract}

\maketitle

\section{Introduction}
Understanding the role that correlations and clustering in nuclear matter and finite nuclei play in determining nuclear structure and dynamic evolution in nuclear collisions has long been a goal of nuclear physics \citeptherein{b1,b2}. In the past three decades recent advances in accelerators, detection systems, theoretical techniques and computing power have fostered major tools to study such phenomena and this has led to important paradigm shifts in our understanding. Rapid progress has been made on many fronts \citeptherein{b3,b4,b5,b6,b7,b8,b9,b10,b11,b12}. Among these, the study of highly excited and/or high spin exotically shaped nuclei, offering the potential of new insights into the nuclear matter equation of state far from stability,  constitutes one of the major focus areas. While light nuclei in their ground states typically exhibit spherical or near-spherical geometries \cite{b13,b14} excited and/or high spin nuclei may exhibit more exotic shapes and density distributions
\cite{b15,b16,b17,b18,b19,b20,b21,b22,b23,b24,b25}. Identifying and examining such nuclei at high excitation is a challenging exercise.

In reference \cite{b26} Cao and collaborators reported detection of high excitation energy resonances in the 7$\alpha$ de-excitation of $^{28}$Si produced in the inverse kinematics reaction 
$^{12}$C($^{28}$Si,7$\alpha$) at 35 MeV/u. The TAMU NIMROD-ISiS 4$\pi$ detection array was employed for that experiment \cite{b27}. A comparison of the reported resonance energies  with those of  theoretical predictions of several groups \cite{b26} led to the suggestion that evidence for long predicted toroidal nuclei \cite{b15,b16,b17,b18} had been observed. No information on the angular momenta of these resonances could be directly extracted from those experimental results. Based on this it was suggested that higher resolution, higher detector granularity, higher statistics experiments be carried out on this and similar systems.

In references \cite{b28,b29} the FAUST \cite{b30} collaboration at TAMU reported results of analyses of such a follow-up experiment. The very thorough data analysis detailed in \cite{b28} shows that a much better resolution and a 20 fold increase in observed events were achieved. However, in the subsequent analysis of this high quality data it was concluded that, the sensitivity of their new measurement “confidently excludes the reported state properties claimed in the previous experiment” \cite{b28,b29}. A key difference in the analyses of the two experiments is the difference in assumptions regarding contributions from un-resolved “background” events and background normalization. Further details of both experiments and their previous analyses may be found in the quoted references \cite{b26,b28}.

Ensuing reviews and additional analyses \cite{b31,b32,b33,b34} reached a different conclusion, i.e.,  that various features of the data reported in references \cite{b28,b29} do, in fact, contain underlying structure. In reference \cite{b31} a comparison of the experimental results to a molecular dynamics calculation assuming disassembly of an excited, rotating, $\alpha$ clustered $^{28}$Si nucleus found structural features similar to those reported in reference \cite{b26}. In \cite{b32} it is reported that an analysis of the 7$\alpha$ events of reference \cite{b28} which include $^8$Be precursors reveals the existence of a number of high excitation energy peaks. The similarity of these excitation energies to those of the recent theoretical calculations of Z. Ren et al \cite{b35} is clear. Determining angular momenta from the existing data is much more difficult and currently relies on model dependent analyses \cite{b31,b32}. In \cite{b34} results of an AI assisted machine learning analysis of both reported 7$\alpha$ excitation energy spectra are presented.  Evidence for regions of very similar, statistically significant, spectral features in the two different data sets was reported. The centroid energies of these regions also indicate correspondence with the predicted energies for $^{28}$Si toroidal resonances \cite{b35}. Motivated by these investigations we have made direct comparisons of the 7$\alpha$ spectra reported in the two different experiments. The results of these comparisons are presented in the following sections.

\section{DATA EVALUATIONS}
\subsection{Spectral Comparisons}
\begin{figure}[htbp]
\includegraphics[width=1.00\hsize]{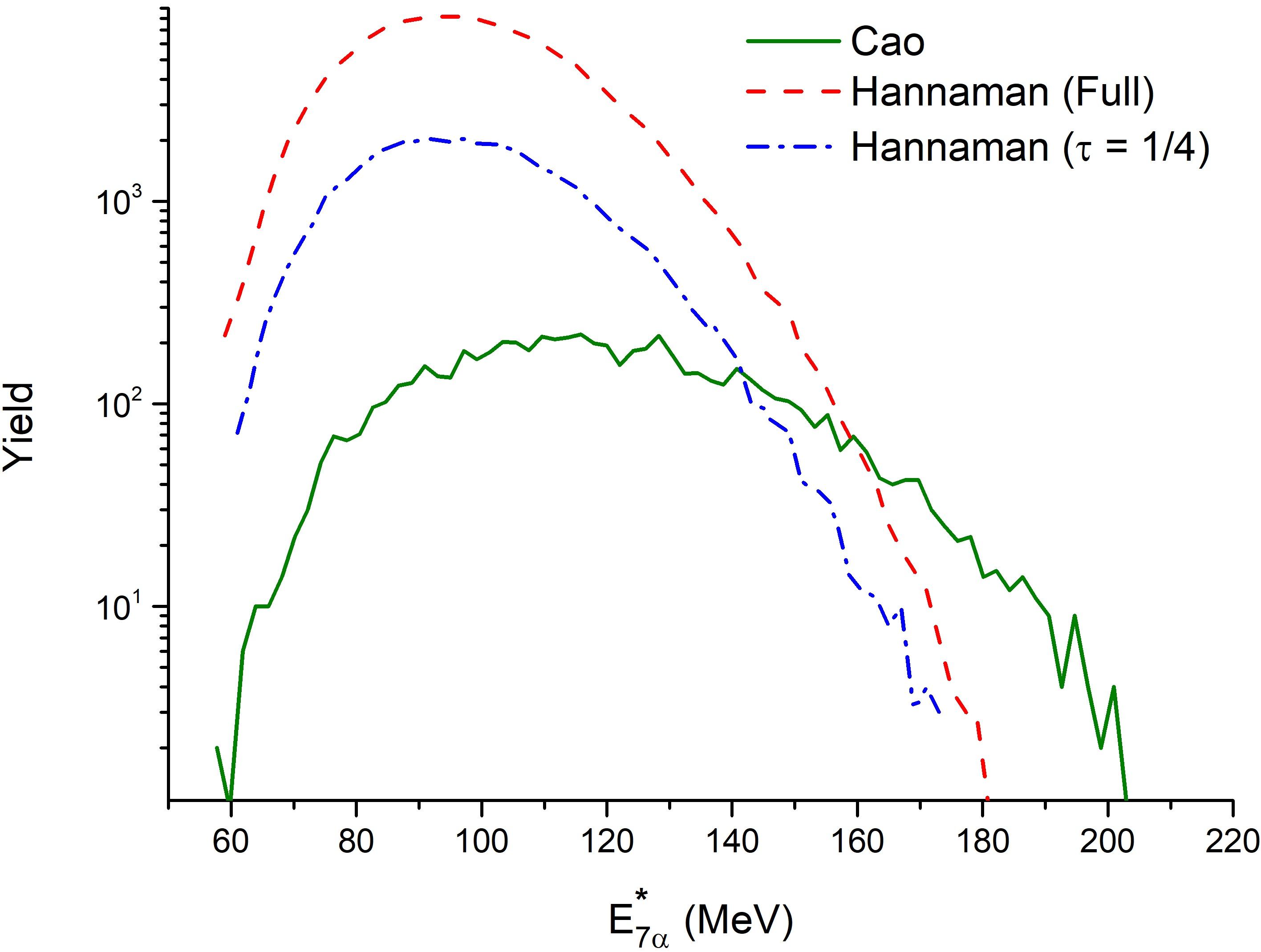}
\caption{Observed 7$\alpha$ excitation energy spectra. Energy bins of 2 MeV are employed. The total data of reference \cite{b28} (top-dashed line) and reference \cite{b26} (bottom-solid line) are depicted. Also shown is the spectrum of events obtained in the  first quarter of the  experimental run of reference \cite{b28} (middle-dash-dotted line). See text.}
\label{fig:Figure1}
\end{figure}

We first present, in Figure \ref{fig:Figure1}, a comparison of the 7$\alpha$ exit channel excitation energy spectra obtained in the two experiments being considered. Several features deserve comment. Immediately observable is that the different experimental acceptances of the 4$\pi$ NIMROD  and the FAUST forward array lead to different observed spectral shapes. As the excitation energy decreases beyond ~90 MeV, the overall statistical advantage of the FAUST experiment decreases.  
The improved statistics of reference \cite{b28}, 180,000 events, is also clearly evident.
To the eye this high statistics spectrum appears much smoother than that of the NIMROD data , 8943 events. This raises the question of statistical significance of apparent oscillations. The much higher statistics of reference \cite{b28} allows further internal spectral comparisons to be made \cite{b33,b34}. In Figure \ref{fig:Figure1} we also present the spectrum of the first 40,000 events obtained in the first part of the experimental run of reference \cite{b28}. This spectrum, already representing a significant increase in statistics over the NIMROD data, has a slightly different shape than the total spectrum. Indeed, time ordered analyses of the data evidences minor fluctuations in excitation energies \cite{b33,b34} which can lead to some smoothing of the  total excitation energy spectrum. Applying a similar process to the much lower statistics Cao results is inconclusive as statistical fluctuations become too large when smaller data samples are isolated.

Even on a logarithmic scale, the two lower curves in Figure \ref{fig:Figure1} exhibit some oscillatory behavior.  Statistical fluctuations are always a concern and the statistical significance of such oscillations requires further exploration. This is addressed in the following discussion.

\subsection{Peak Searching}
The use of spectral derivatives to search for peaks is a standard technique \cite{b36,b37}. In reference 
\cite{b31} spectral derivatives of the data of Cao et al. were employed to search for peaks. This exercise supported the claims of structure reported in \cite{b26} and indicated an existence of lower energy structures as well. Given the low statistics of the data no further investigation was undertaken at that time. In the present work we employ second derivatives to make a direct comparison between the spectra from the two different experiments. We emphasize that this approach is sensitive to the different statistical and systematic errors of the two experiments, but is free of any background assumptions. 

\begin{figure}[htb]
\includegraphics[width=1.00\hsize]{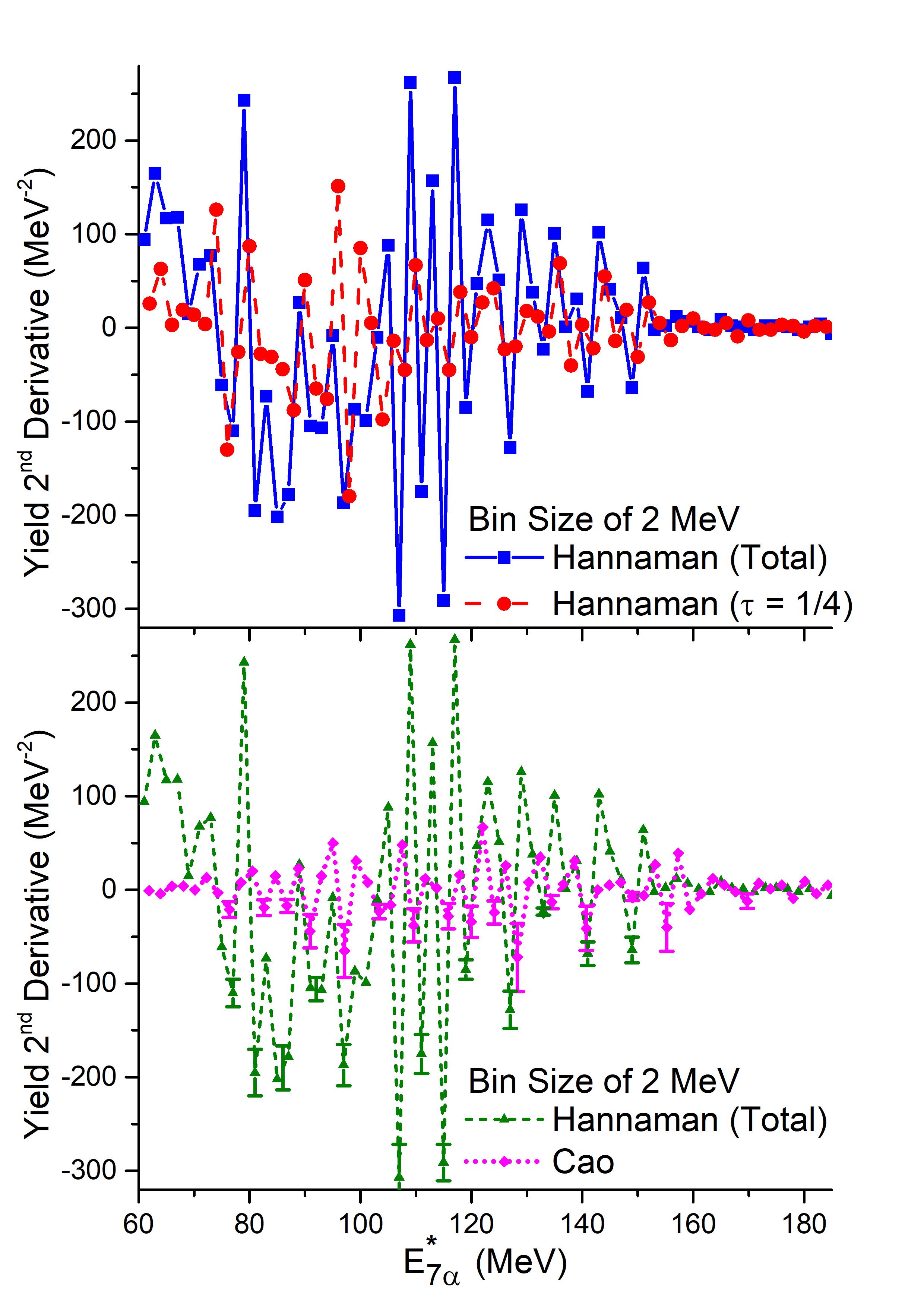}
\caption{Second derivative analyses of the available $^{28}$Si \textrightarrow 7$\alpha$ Data. Top- Comparison of derivatives for data from FAUST experiment \cite{b28} Top-Total-solid squares, first $\sfrac{1}{4}$ of data-solid circles. Bottom- Comparison of derivatives for data from FAUST \cite{b28}, triangles, and NIMROD, diamonds \cite{b26}. Error bars indicate uncertainties on amplitudes of negative minima. See text.}
\label{fig:Figure2}
\end{figure}

In our comparisons of second derivatives of samples of the time ordered data of reference  \cite{b28} we noted small variations reflective of the spectral shape difference already observed in Figure \ref{fig:Figure1}. This is clearly depicted in the top portion of Figure \ref{fig:Figure2} where the second derivative of the total 180,000 event spectrum is compared to that of first one quarter of those events (in time). The two results are very similar, but generally offset by $\sim$2 MeV in the locations of the observed minima. Some small shape differences, resulting from either statistical fluctuations or possible fluctuations in responses of individual detectors are also seen.
In the initial comparison of second derivatives for the two reported 7$\alpha$ spectra reported 
\cite{b26,b28} we observed a very small difference in energy scales and peak positions indicating a very small systematic difference in the energy calibrations of the two different experiments. To improve the comparison we chose to remove this systematic difference by making   a slight adjustment of the energy calibration from the NIMROD experiment \cite{b26} to better align it with the higher statistics, higher resolution data of the FAUST experiment \cite{b28}. The difference between the newly adopted calibration and that used previously increases with excitation energy. This leads to an excitation energy dependent increase of $\sim$ 2\% in the excitation energies reported in \cite{b26}.The energies of the three peaks originally reported in reference \cite{b26} increase to 116 MeV, 128 MeV and 141 MeV.

Our final comparison of second derivatives of the total 7$\alpha$ spectra reported for the two experiments is then presented in the bottom of Figure \ref{fig:Figure2}. The statistical advantage at lower excitations and the better resolution of the data from reference \cite{b28} are apparent.  More importantly, it can be seen that, even with the different statistical and systematic uncertainties inherent in the two experiments, the second derivatives of both data sets exhibit negative minima, with very significant similarities in the excitation energies at which these second derivative minima can be seen. The first five columns of Table \ref{tab1} contain a summary of identified 7$\alpha$ producing state energies extracted from Figure \ref{fig:Figure2} and, together with the excitation energies derived in \cite{b32}, the toroidal states predicted by Ren et al. \cite{b35} and the AI assisted spectral regions identified by Depastas et al. \cite{b34}.  Earlier theoretical predictions for toroidal states can be found in \cite{b18,b26}.

\begin{table*}[!htbp]
\caption{Identified 7$\alpha$ producing states in excited $^{28}$Si. All energies are in MeV.}
\label{tab1}
    \centering
\begin{tabular}{|c|c|c|c|c|c|}
\hline

\textbf{Hannaman \cite{b28}} & \textbf{Cao \cite{b26}} & \textbf{WADA \cite{b32}} & \textbf{Ren Predictions \cite{b35}} & \makecell{\textbf{Depastas Al} \\ \textbf{Region Centroids \cite{b34}}} & \makecell{\textbf{Hannaman 95\%}\\\textbf{Confidence level}}\\

\hline
& & 69.5 $\pm$~0.4 & & 63.8 $\pm$~2.4 & 67 $\pm$~1\\
\hline
77 $\pm$~1.2 & 76 $\pm$~1.2 & 76 $\pm$~0.4 & 72.7 & 73.1 $\pm$~0.88 & 75 $\pm$~1\\
\hline
81 $\pm$~1.2 & 82.6 $\pm$~1.2 & 81.5 $\pm$~0.4 & 81.6, 81.8 & & 83 $\pm$~1\\
\hline
86 $\pm$~1.2 & 86.7 $\pm$~1.2 & 86.5 $\pm$~0.4 & & 85.1 $\pm$~1.6 &\\
\hline
92 $\pm$~1.2 & 90.9 $\pm$~1.2 & 93.5 $\pm$~0.4 & 89.6 & & 91 $\pm$~1\\
\hline
97 $\pm$~1.2 & 97.1 $\pm$~1.2 & 100.5 $\pm$~0.4 & 98.3, 98.5 & 101 $\pm$~2.5 & 101 $\pm$~1\\
\hline
107 $\pm$~1.2 & 103 $\pm$~1.2 & 105.5 $\pm$~0.4 & 106.3 & &\\
\hline
111 $\pm$~1.2 & 110 $\pm$~1.2 & 110.5 $\pm$~0.4 & & & 109 $\pm$~1\\
\hline
115 $\pm$~1.2 & 116 $\pm$~1.2 & 114.5 $\pm$~0.4 & & & 117 $\pm$~1\\
\hline
119 $\pm$~1.2 & 120 $\pm$~1.2 & 118.5 $\pm$~0.4 & & &\\
\hline
 & 124 $\pm$~1.2 & 122.5 $\pm$~0.4 & & 121 $\pm$~4.11 &\\
\hline
127 $\pm$~1.2 & 128 $\pm$~1.2 & 127 $\pm$~0.4 & 128, 128.1 & & 127 $\pm$~1\\
\hline
133 $\pm$~1.2 & 134 $\pm$~1.2 & 134 $\pm$~0.4 & & &\\
\hline
141 $\pm$~1.2 & 141 $\pm$~1.2 & 140.5 $\pm$~0.4 & & 139 $\pm$~4.5 &\\
\hline
149 $\pm$~1.2 & 149 $\pm$~1.2 & & 147.9 & &\\
\hline
 & 155 $\pm$~1.2 & & & &\\
\hline
 & 170 $\pm$~1.2 & & 168, 168.2 & & 169 $\pm$~1\\
\hline
Neg 2nd Der. & Neg 2nd Der. & $^{8}$Be filter & & &\\
\hline
\end{tabular}
\end{table*}

While the lists in Table \ref{tab1} are not identical, within uncertainties of a few MeV in the total excitation energies, there is strong agreement among the values derived from the two experiments \cite{b26,b32,b33}. Also observable is that a number of the identified states have energies close to those of the toroidal states predicted in reference \cite{b35}. While direct information on the actual geometries of and the spins of the identified states are not currently accessible, the agreement of the excitation energies of many of them with the toroidal state predictions provides a strong circumstantial case that toroidal states have been observed. The unsupervised machine learning Gaussian Mixture Model (GMM) model employed in \cite{b34}, applied to the data from \cite{b28}, identified 6 Gaussian regions with statistically significant structures. The centroid values of those Gaussians, presented in the fifth column of Table \ref{tab1}, are correlated with the locations of the individually identified states. A GMM analysis with a larger component number is part of a later section in this paper.

Additional peaks not corresponding to predicted toroid states have also been identified. Of these, some might arise from statistical fluctuations, or incomplete detection of de-excitation products from higher mass nuclei. Additional favored geometries might be indicated. 

\begin{figure}[tbp]
\includegraphics[width=1.0\hsize]{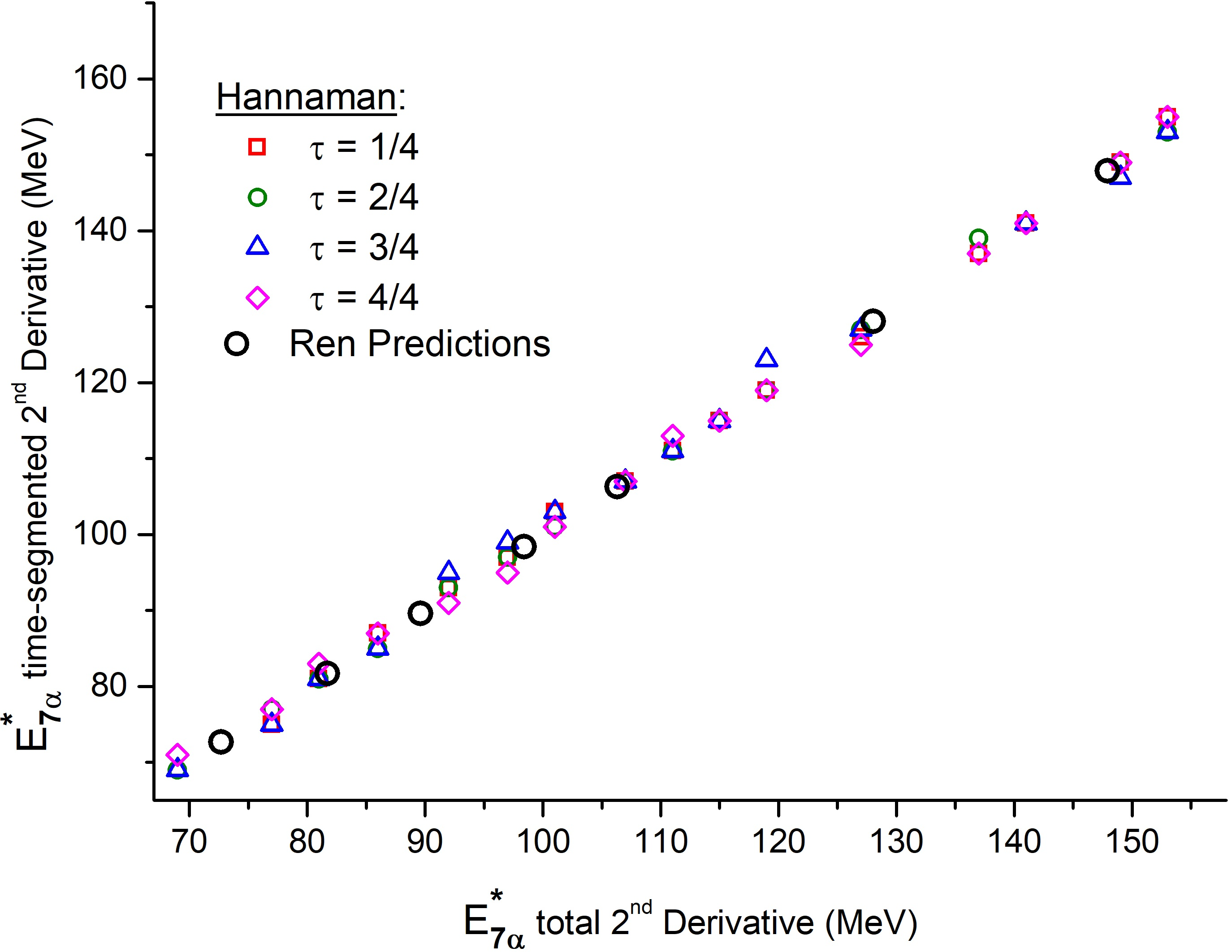}
\caption{Excitation energies of second derivative derived peaks for four separate time segments vs excitation energies derived from the total 7$\alpha$ spectrum. Also shown are the predicted toroidal energies \cite{b35} {plotted against themselves}. See text.}
\label{fig:Figure3}
\end{figure}

To further elucidate the second derivative results we present in Figure \ref{fig:Figure3} a summary plot comparing the second derivative minima derived from the 4 sequential (in time) quarters of the FAUST data to the Ren et al, lattice DFT calculations for toroids \cite{b35}.

That there are a number of peaks corresponding closely to the toroid predictions of \cite{b35} is again apparent. Other peaks are also identified, While some of the smaller statistics peaks might reflect statistical fluctuations, their being observed in multiple segments favors their existence and warrants further investigation. Predictions of toroidal and linear states in other highly excited nuclei exist \cite{b18,b40}. In particular we note that stabilized linear states are predicted for excitations of ~80 to 100 MeV $^{28}$Si and angular momenta from 16 to 36  $\hbar$ are expected for $^{28}$Si \cite{b40}. Note that, not only the energies, but also the angular momenta of these overlap those predicted for the lower excitation toroidal configurations \cite{b35}.

Finally, the FAUST experiment also obtained a sufficient number of 8$\alpha$ events to allow a second derivative analysis similar to that performed for the 7$\alpha$ data.  While some $^{32}$S nuclei can be produced in $\alpha$ pick up reactions, that  analysis of the in frame $\alpha$ particle velocity distributions \cite{b28} demonstrates clearly that the bulk of the  detected  8$\alpha$ events  correspond to events in which the 8th $\alpha$ is emitted in breakup of $^{12}$C target nuclei and has  high energy in the $^{28}$Si projectile frame. Such events were rejected in the NIMROD experiment.

Interestingly, as seen in \cite{b28}, the reported 8$\alpha$ excitation energy spectrum appears to have more structure than the reported 7$\alpha$ spectrum. This suggests that requiring the simultaneous detection of break-up products from the target produces a filtering effect which reduces background contributions, possibly by narrowing of the deposited angular momentum window for the selected exit channels. Removal of that trailing 8th $\alpha$ particle on an event by event basis and analysis of the remaining projectile-like contribution should allow a direct comparison with the analysis of the events identified as 7$\alpha$ presented in Figure \ref{fig:Figure2}.

\begin{figure}[tbp]
\includegraphics[width=1.0\hsize]{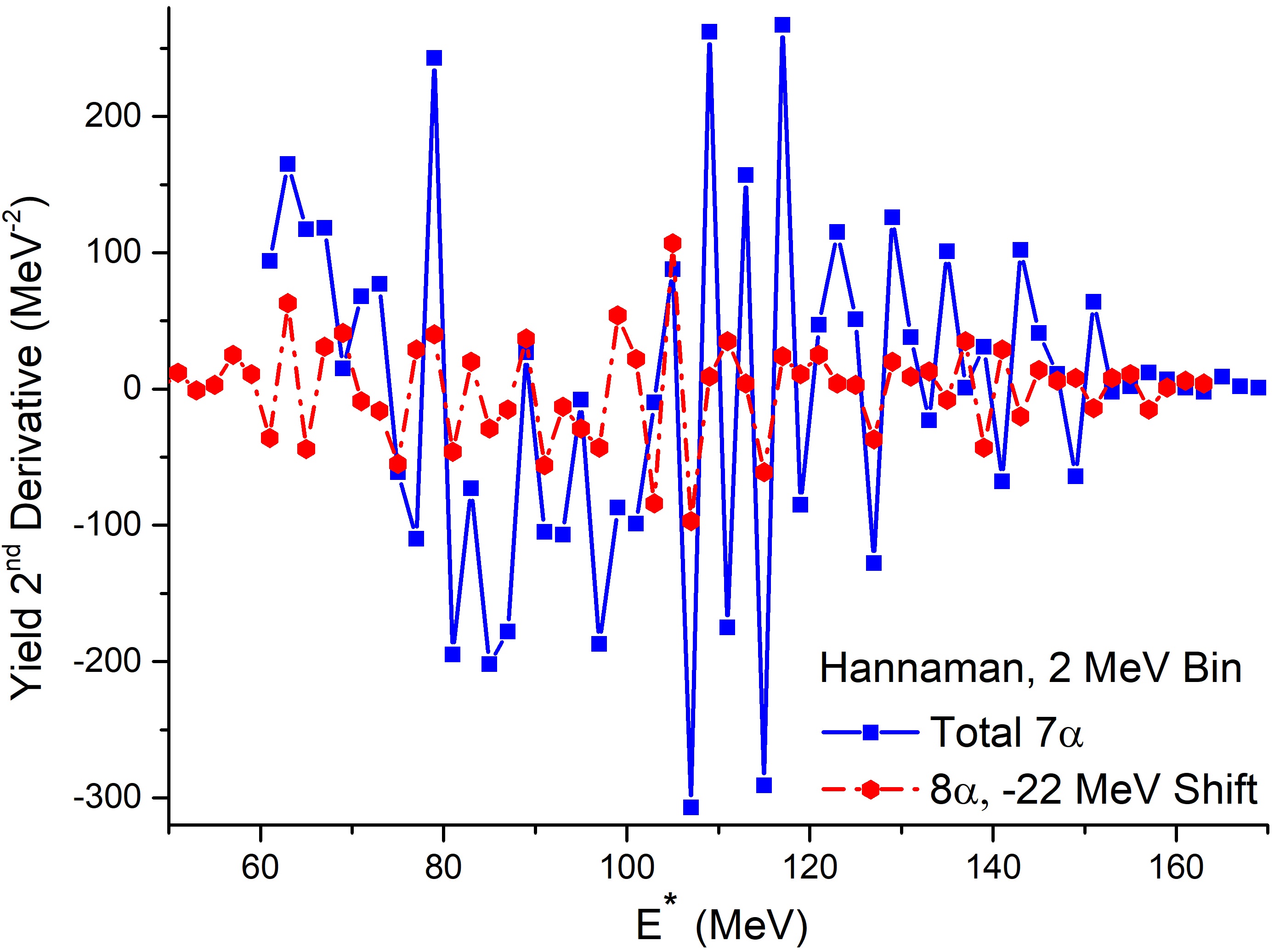}
\caption{Second derivatives 7$\alpha$, solid squares, and 8$\alpha$ (shifted), solid circles, FAUST Events \cite{b28}.}
\label{fig:Figure4}
\end{figure}

Another approach is recognition that plots of the reported 7$\alpha$ and 8$\alpha$ excitation energy spectra \cite{b28} show that the centroid of the 8 $\alpha$ spectrum is 22 MeV higher than that of the 7$\alpha$ spectrum. This results from the inclusion of the kinetic energy of the $\alpha$ particle from $^{12}$C and the Q value difference for the $\alpha$ particle disassembly of $^{32}$S and $^{28}$Si. On average, shifting the 8$\alpha$ spectrum downward 22 MeV in excitation energy then allows a comparison to the results shown in Figure \ref{fig:Figure2}. That comparison of second derivatives is shown in Figure \ref{fig:Figure4}. It reveals a considerable similarity, confirming the near identical structural features of the two spectra. That similarity is detailed in Table \ref{tab2}, in which the locations of negative minima in the second derivatives are recorded.

\begin{table}[!htb]
\caption{Second derivative negative minima from Figure \ref{fig:Figure4}. All energies are in MeV.}
\label{tab2}
    \centering
\begin{tabular}{|c|c|} 
\hline
7$\alpha$ min2ndDer & 8$\alpha$ min2ndDer(EX-22 MeV)\\ [0.5ex]

\hline
$77$ & $75$ \\ 
\hline
$81$ & $81$ \\ 
\hline
$86$ & $86$ \\ 
\hline
$92$ & $91$ \\ 
\hline
$97$ & $97$ \\ 
\hline
$107$ & $107$ \\ 
\hline
$111$ & $ $ \\ 
\hline
$115$ & $115$ \\ 
\hline
$119$ & $ $ \\ 
\hline
$127$ & $127$ \\ 
\hline
$133$ & $135$ \\ 
\hline
$141$ & $143$ \\ 
\hline
\end{tabular}
\end{table}

The 111 MeV, and 119 MeV peaks observed in the 7$\alpha$ column are not predicted as toroidal states and are missing from the  8$\alpha$ column.
As already suggested above they might result from other exotic $\alpha$ clustered shapes, e.g., linear, prolate, bubble which might be produced in the early stages of the collisions \cite{b16,b24,b38,b39,b40,b41,b42}. Experiments investigating  $^{32}$S and other nuclei would be useful in resolving such questions as would experiments employing different probes and analysis techniques \cite{b41,b43,b44}. As previously pointed out \cite{b15,b16,b26,b32,b33} instabilities of excited N$\alpha$ clustered structures are unlikely to decay simultaneously into N $\alpha$s so that more detailed information on the de-excitation sequence should be sought.

\subsection{Statistical Significance}
In a previous version of this paper we proposed that the high level of agreement of the background free second derivatives for the two different experiments with different systematic and statistical errors, as well as the obvious overlap of the 7$\alpha$ and 8$\alpha$ results from \cite{b28,b29} argued strongly for real underlying structure in the 7$\alpha$ exit channel data. As statistical fluctuations are always with us, the referee of our paper requested that we carry out a more complete evaluation exploring the role that statistical fluctuations could be playing in the data. Based upon that suggestion, and aided by new AI tools, we have carried out a series of analyses focused on better characterizing the data through use of well-established statistical techniques. Specifically we have employed Savitsky-Golay filtering with Hampel filtering \cite{b45,b46} verification and GMM analysis combined with multi‐scale wavelet decomposition with probabilistic and Bayesian inference-while controlling for minor calibration drifts and employing multiple comparisons to discriminate against random statistical resonances. Our results are summarized in this section.

Data and Pre-processing

Employing 2 MeV bins for all data, we first computed the second derivative of each spectrum analyzed to locate candidate curvature minima. Light Savitzky–Golay smoothing \cite{b45} (window = 7, order = 3) preserved peak shapes while reducing noise. Applying Hampel filtering \cite{b46} (window = 7, threshold = 3 × Medium Absolute Deviations) then flagged robust local maxima for preliminary validation. A Daubechies‐4 discrete wavelet transform \cite{b47} was then applied to to each spectrum, reconstructing a level-4 approximation as the broad baseline. This revealed a concealed hump peaking near 90–92 MeV with half‐width of 75–85 MeV. Subtraction of this baseline produced detailed spectra that isolated narrow resonances without manual background fitting.

Following that, again with AI tools, we addressed peak detection and validation according to the following sequence.

1)	Savitzky–Golay + Second Derivative
\begin{itemize}[label=$\circ$, nosep]
\item Located candidate peaks at derivative minima.
\end{itemize}

2)	Hampel Filter
\begin{itemize}[label=$\circ$, nosep]
\item Confirmed candidates exceeding local 3 $\times$ Medium Average  Deviation thresholds.
\end{itemize}

3)	Gaussian Mixture Model with Hybrid Background
\begin{itemize}[label=$\circ$, nosep]
\item Initialized Gaussians at derivative minima combined with a broad background component for which statistically insignificant candidate peaks
were removed using Bayesian Information and Akaike Information criteria and extracted peak areas and uncertainties.
\item Declared peaks significant for Signal to Noise Ratios = area/$\sigma_{area}$ $\geq$ 3 and Full Widths at Half Maxima  $\geq$ 3 MeV.
\end{itemize}

4)	Bayesian Matched‐Filter with False Discovery Rate Control \cite{b48}
\begin{itemize}[label=$\circ$, nosep]
\item Convolved detail spectra with Voigt templates (widths 3–6 MeV).
\item Fitted peaks plus local linear background under Poisson likelihood.
\item Sampled posteriors via Markov Chain Monte Carlo, demanded posterior probability amplitudes >0) and applied the Benjamini–Hochberg procedure \cite{b49} to control false discovery rate at 5\%.
\end{itemize}

For the comparison of five different spectra, i.e., of the 4 time-ordered segments and the total sum spectrum of the high statistics data of reference \cite{b28,b29} the wavelet‐corrected Bayesian and Gaussian Mixture Model analyses yield identical core resonance lists across all five spectra: These resonances, at the 95\% confidence level, are at 67 MeV, 75 MeV, 83 MeV, 91 MeV, 101 MeV, 109 MeV, 117 MeV, 127 MeV and 169 MeV. The last column of Table \ref{tab1} lists these. Evidence for structures at 95 MeV, 97 MeV, 137 MeV, 145 MeV, 157 MeV and 159 MeV was determined to be marginal and these were rejected.

For the much lower statistics data of reference \cite{b26} larger statistical fluctuations can lead to additional possible peaks. Therefore, for the statistical analysis of that data we constrained the full widths at half maximum of the allowed peaks to be greater than or equal to 3.3 MeV, in the range consistent with the widths obtained from the analyses of the much higher statistics experiment. With that constraint, robust peaks were found at 83 MeV, 91 MeV, 101 MeV, 117 MeV, 127 MeV and 165 MeV. Thus a significant degree of agreement remains for the two different experiments.

It is worth noting that if resolution is $\sim$ 2 MeV and observed widths are $\sim$ 3.3 MeV and as natural widths, lifetimes near 50 fm/c are implied. Such lifetimes would be consistent with early dynamic formation and fragmentation. Also, assuming the estimated total 7$\alpha$ cross section of reference \cite{b26}, individual resonance cross sections can be estimated from the results of this analysis. These range from $\sim$ 35 µb at the peak of the 7$\alpha$ excitation function to $\sim$ 10 µb at the higher and lower excitation energies. Approximately 25\% of the 7$\alpha$ cross section is found in the observed resonances. The remainder constitutes the background.

Applying the same AI analysis to the data of reference \cite{b26} in 2 MeV histogram bins, and slightly energy shifted as described in Section B, indicated confirmed resonances at 103, 116 and 130 MeV.

Carrying out such an analysis on 100,000 cases of sets of four simulated raw energy spectra containing only random fluctuations, we found that the probability of four such individual random spectra finding overlap probabilities of the same 2 energies was 19.3\%, the same 6 energies was 8.3 $\times$ 10$^{-2}$\% and the same 9 energies was 6 $\times$ 10$^{-4}$\%. This last number is orders of magnitude below that observed for the four time-ordered segments of the high statistics Hannaman data.

\section{Conclusions}
A second derivative analysis of the currently available $^{12}$C($^{28}$Si,7$\alpha$) data reveals high excitation energy peak structures in two reported 7$\alpha$ excitation energy spectra \cite{b26,b28} as well as in the 8$\alpha$ spectrum of Ref. \cite{b28}. The analysis makes no a priori assumption regarding possible background events contributing to these spectra. Strong similarities in the peak energies derived in two experiments employing different detectors are observed. These peak energies are generally well aligned with those reported in \cite{b32} and with the analysis of \cite{b34}. Such agreement argues strongly against statistical fluctuations as their origin. 
A detailed statistical filtering analysis employing well-established, powerful techniques confirms the existence of real structural features in all spectra examined.
The close agreement of many of these states with excitation energies predicted for stabilized toroidal nuclei is noted and the need for further information, in particular on spins, is emphasized. Other favored geometries may also exist, as discussed in the literature \cite{b16,b24,b38,b39,b40,b41,b42}.

Determination of deposited angular momenta would considerably facilitate direct comparison of data with the theoretical predictions. While energy deposition should increase (with significant fluctuations) as the reaction impact parameter decreases below that of the grazing collision, the angular momentum deposition should first increase and then decrease (again with large fluctuations) as the impact parameter approaches zero.  This indicates that the simultaneous deposit of the correct excitation energy-angular momentum combinations predicted for the toroidal states should be dominated by mid-peripheral collisions. Comparison of the data with reaction models can provide simultaneous estimates of energy and angular momentum depositions which can be instructive \cite{b32}. More detailed cluster correlation measurements may also provide angular momentum information \cite{b42,b43}.

To this point, searches for toroidal structures in the data from the two experiments have been focused upon the 7$\alpha$ data. If toroidal, or other exotically shaped nuclei with high $\alpha$ localization coefficients \cite{b35,b40} are responsible for the experimentally observed resonances, the other $\alpha$-conjugate exit channels, spanning a wider range of excitation energies, can be examined for analogous structural features. Such additional experimental and theoretical explorations of collisions of $\alpha$ conjugate nuclei \cite{b31,b34} can provide new insights into $\alpha$ clustered matter.

Also, $\alpha$ clustered states may act as doorway states, some of which de-excite quickly into $\alpha$ conjugate exit channels, while others merge into the continuum and de-excite by other pathways. If so, certain non-$\alpha$-conjugate exit channels may exhibit enhancements at the doorway state resonance energies. Our analyses of additional data from $^{28}$Si + $^{12}$C reactions are currently being extended to explore this possibility. At very low impact parameters the dynamic production of toroidal structures oriented perpendicular to the beam direction may also be observable \cite{b50,b51,b52}. 

\section{Acknowledgements}
JBN thanks A. Hannaman for many open and valuable interactions, C. Y. Wong for his many contributions to my understanding over many years of discussion, J. Meng and Z. Ren for discussion and sharing some unpublished results, L. Sobotka for his incisive comments and questions as this work was in progress and the myriad AI developers for facilitating the statistical analysis. This work was supported in part by the United States Department of Energy under Grant \#DE-FG02-93ER40773.

\bibliographystyle{unsrt}

\end{document}